# A Review of Artificial Intelligence Impacting Statistical Process Monitoring and Future Directions

Shing I Chang and Parviz Ghafariasl
Industrial and Manufacturing Systems Engineering
Kansas State University, Manhattan, Kansas
Corresponding author: changs@ksu.edu

## Abstract

It has been 100 years since statistical process control (SPC) or statistical process monitoring (SPM) was first introduced for production processes and later applied to service, healthcare, and other industries. The techniques applied to SPM applications are mostly statistically oriented. Recent advances in Artificial Intelligence (AI) have reinvigorated the imagination of adopting AI for SPM applications. This manuscript begins with a concise review of the historical development of the statistically based SPM methods. Next, this manuscript explores AI and Machine Learning (ML) algorithms and methods applied in various SPM applications, addressing quality characteristics of univariate, multivariate, profile, and image. These AI methods can be classified into the following categories: classification, pattern recognition, time series applications, and generative AI. Specifically, different kinds of neural networks, such as artificial neural networks (ANN), convolutional neural networks (CNN), recurrent neural networks (RNN), and generative adversarial networks (GAN), are among the most implemented AI methods impacting SPM. Finally, this manuscript outlines a couple of future directions that harness the potential of the Large Multimodal Model (LMM) for advancing SPM research and applications in complex systems. The ultimate objective is to transform statistical process monitoring (SPM) into *smart process control* (*SMPC*), where corrective actions are autonomously implemented to either prevent quality issues or restore process performance.

## Introduction

Water Shewhart first converted a series of hypothesis tests into a graphical form for AT&T production facility in 1924 (W. Shewhart 1931). Chat-GPT(OpenAI 2024) summaries Shewhart's contribution as follows:

> "Shewhart's key insight was that variation in manufacturing processes could be categorized into two types: common cause variation and special cause variation. Common cause variation arises from the inherent variability of a process and is expected to occur even under stable conditions. On the other hand, special cause variation occurs due to specific factors that are not inherent to the process and can be identified and eliminated."

The accuracy and quality of this summary is impressive. It captures the essence of how control charts intend to work in distinguishing common cause and special cause variation. Other recent Artificial Intelligence (AI) advances include Convolutional Neural Networks (CNN) (LeCun et al. 1998) and RNN (Rumelhart, Hinton, and Williams 1986). CNN is often used for imaging related applications. CNN has been widely adopted for visual inspections when images can be directly fed into a trained model for real-time inspections. RNN is a model capable of learning from data history. Several CNN models trained from data from different time periods can be stringed together using the RNN structure tackling various time-series problems. The AI related advances have sparked the imagination of how AI technologies can be applied to solve complex SPC problems, which are dominated by traditional statistical methods. For example, an X-bar chart uses sample means for detecting process mean shifts while an R chart or S chart uses sample range or standard deviation for monitoring process variation changes. Both charts combined are intended for process monitoring. If all plotting statistics, i.e., sample means, ranges, or standard deviations, fall within the control limits without obvious patterns, then the process under surveillance is deemed in control. Otherwise, the process may be out of control due to assignable causes. This type of control charting is a classic binary classification problem that existing AI/ML models can solve. With the advancement of AI technology, such as generative AI via the transformer mechanism (Vaswani et al. 2017), we expect AI to solve more complex problems, such as system-wide process monitoring for semiconductor production.

The era of the digital revolution has arrived. The German-led Industrial 4.0 (Hermann, Pentek, and Otto 2016) articulates the next generation of manufacturing, where data generated from the entire production ecosystem from supply chains, production equipment, in-process production data, inspection data, and customer feedback are all captured and digitized. Digital twins, a virtual factory, can be built based on this vision. Hypothetical changes in virtual factories will not affect the physical production, allowing optimized operations to be planned in a virtual environment (Grieves 2014). In addition, digital records contain all data in both spatial and temporal dimensions. In other words, it includes records from everywhere in a system and from the past to the present. Would it be possible to build a virtual SPC system that foresees pending quality problems rather than detects existing problems in a production ecosystem? Would it be plausible to fine-tune upstream process parameters to prevent pending downstream quality issues? AI-based approaches rather than statistics-based modeling approaches will have a better chance to provide an answer to this vision.

This manuscript first provides a brief introduction to statistics-based control charting methods. The scope is limited to control charting for process monitoring. This manuscript does not include tools often used for continuous process improvement, such as cause-effect diagrams, flowcharts, check sheets, …, etc. Both fault detection and anomaly detection are also reviewed, but the focus is on processes rather than machinery. Then, a taxonomy is provided to classify various AI methods, including neural networks (NN) and machine learning (ML) developed for SPC applications over the last 25 years. We will compare these two schools of thought regarding performance metrics, setup or training methods, and the possibility of collaborative use. Several AI modeling approaches will be explored to spot the trend for future SPC/SPM research and applications. Finally, we will articulate two future directions of leveraging generative AI for achieving system-wise monitoring and smart process control.

# A Brief History of Statistics-Based Control Charting

**Review of Univariate Control Charts**

The principle of SPC emphasizes that process adjustment is needed only when assignable causes for variations are detected. Common-cause variations occur in any production process. As long as variations in part dimensions stay within the "boundaries," control actions will not be triggered. SPM is a preferred term over SPC because control charts do not provide any control to restore an out-of-control process to the in-control state. It merely informs operators that the process may be out of control and control actions may be required. The responsibility of control actions falls sorely on operators or process engineers.

Attribute control charts and variable control charts are the two main SPM categories depending on the data types of the quality characteristics, often on products or semi-finished workpieces. For single discrete quality characteristics, Binomial and Poisson distributions are the most popular statistical models used for attribute control charts, such as the P chart (i.e., fractional nonconformity control chart) and c chart (i.e., nonconformity control chart). Variable control charts such as X-bar, R charts (Shewhart 1930), exponential weighted moving average (EWMA) charts (Roberts 1959), and cumulative sum (CUSUM) charts (Page 1954) have been used for process control where its quality characteristic is a continuous variable, such as a weight or a dimension. X-bar charts are capable of detecting large process shifts, while EWMA and CUSUM are suitable for small process shifts. In addition, both EWMA and CUSUM are effective when individual observations are inputs, while X-bar and R charts rely on sample sizes (n) larger than one. EWMA charts assigned more weights on recent data points, while the CUSUM charts assigned equal weights on all data points. Central Limit Theorem (CLT) states that the sample means are Normally distributed if the observations forming a sample are homogeneously distributed even if n is small (e.g., n= 4 or 5). CLT is the main reason that X-bar and R charts are robust and still widely used today.

**Review of SPM When Data Assumptions Are Violated or Historical Data is Scarce**
Although CLT can ensure samples of the non-Normal observations approximate Normal distributed data. Other data assumption violations warrant attention for control charts to operate properly. These violations include correlated process observations and insufficient setup data for job-shop productions (Lin, Lai, and Chang 1997). When SPM is applied to process parameters, observations may not be statistically independent. Several approaches to address this challenge include down sampling, residual charting based on EWMA forecasting, and using time-series models. When process observations are collected quickly, adjacent observations will exhibit autocorrelation. A quick remedy is to down sample by keeping only one out to N (e.g., 10) observations to break autocorrelation. However, this strategy increases the reaction time by N fold and may not always work. An improved method is to use all observations in EWMA statistics. Then the residuals are generated from the differences between the observation at time t and the EWMA value at time t-1 (i.e., after the observation at time t-1 is computed). If the EWMA model can adequately mimic the autocorrelation behavior. The residuals should be Normally distributed with constant standard deviation. In the cases where the performance of the EWMA model is lacking, more general time series models, such as autoregressive integrated moving averages (ARIMA), should be considered (Jenkins and Box 1976). The residuals are differences between the observed and predicted data and are fed into control charts. Note that it can be shown that EWMA is a special case of the ARIMA model (i.e., ARIMA (0,1,1) model with moving average coefficient $\theta$) when the EWMA parameter $\lambda$ is chosen properly i.e., $\lambda=1-\theta$, (Montgomery, Jennings, and Kulahci 2015).

A standard guideline for control chart setup requires sufficient data, e.g., 100-125 observations. When this condition is not met or the preference is to start a control chart immediately, several mitigation strategies include self-start control charts, short-run SPC, and transfer learning. Hawkins (1987) proposed a self-starting CUSUM chart for location and scale. The proposed method aims to monitor whether there have been drifts from the conditions obtained at the process startup, although the process mean and standard deviation are unknown initially. The core idea is to use a pair of CUSUM charts – one for mean monitoring and the other for spread monitoring. Running mean and standard deviation estimates substitute the actual process parameters. As more and more observations are collected, the performance of the charts will converge when sufficient historical data is available.

Short-run control charts are often adopted when a production process only produces a few items of the same kind and then switches to another product. Assuming the variation of the process remains the same for all products produced on a process, the short-run SPM method first transforms the plotting statistic by subtracting the observed value from its nominal or target value. Then, the transformed value is plotted on the same control chart for multiple products using the same process. The transformed data collected from various products may be pooled to estimate the control limits. In short-run situations, individual charts are often used, and the moving range statistics can also be pooled for process standard deviation estimation. Lin, Lai, and Chang (1997) proposed a part family formation method to group products produced on the same process based on the process standard deviation estimates. In this case, products grouped in the same family will be plotted on the same chart. If one control chart is preferred, then the plotting statistic $z_i$ is standardized as: $z_i = (x_i - T_j)/s_j$ where $T_j$ and $s_j$ are the target value and standard deviation of the *jth* group, respectively. Finally, there are situations where a new product is introduced without any production history. If prior productions contain data from similar products, then the knowledge can be gleaned to start the process monitoring of the new product. In some regards, this approach is very similar to the part-family strategy. For more detail on this subject, Tsung et al. (2018) conducted a literature review regarding both statistical models and methodologies on statistical transfer learning. When a new part enters production, data from similar previously produced parts is used to estimate control limits for the new part.

**Review of Non-Control-Chart Tool for SPC**
As opposed to real-time, online process monitoring, process analysis is a study of process capability conducted after process data has been collected. The results of this analysis can be used either to plan for a control chart on a new product or to provide quality assurance information to satisfy regulatory requirements, such as ISO 9000 standards (Sun 2000). Typically, a single value provided by a process capability index (CPI) summarizes the information about process variability related to the specification of its quality characteristic. One of the most popular CPIs is simply the ratio of the range of specification limits to six sigma (process standard deviation)(Truscott 2012).

One of the key challenges in non-control-chart-based SPC tools is selecting the appropriate criterion for different process conditions. In automated and AI-driven manufacturing, sudden and subtle changes in process variables can impact product quality. Traditional Statistical Process Control (SPC) methods, such as process capability indices (Cpk, Ppk), rely on restrictive assumptions and fail to provide an accurate representation of process dynamics (Montgomery 2020). Hence, more advanced analytical techniques are required for effective process monitoring. Modern SPC methods have been developed to enable more precise analysis of production data. Wavelet-based monitoring identifies local anomalies without assuming normality, making it useful in industries like semiconductors and pharmaceuticals (Bakshi 1998). Time series decomposition techniques, such as ARIMA, GARCH, and Kalman filters, help detect short- and long-term trends and are widely used in automotive, petrochemical, and aerospace industries (Hamilton 1994; Box et al. 2015). Additionally, Functional Data Analysis (FDA) is valuable in industries where process data is continuously recorded over time, such as steel manufacturing and food processing, by uncovering hidden patterns in process behavior (Ramsay and Silverman 2005). Kotz and Johnson (2002) provided a summary of both univariate and multivariate process capability indices. However, critics contend that merely using a single number to represent the entire history of a time series data was fundamentally flawed because the index itself was also a random variable subject to the influence of variation.

## Review of SPM Based on Control Chart Patterns

Control chart users primarily rely on three-sigma control limits to determine whether a process is in control. However, patterns formed by a sequence of points on a control chart can also indicate potential out-of-control conditions. The Western Electric Rules (1958) defined several key indicators, including one or more points outside the control limits, two out of three consecutive points beyond the two-sigma limits, four out of five beyond the one-sigma limits, and eight consecutive points on one side of the centerline. Additional indicators included trending patterns, cyclic behaviors, and abrupt shifts. While these rules enhance process anomaly detection, increasing the number of detection rules raised the risk of false alarms, making operator judgment essential. Over time, several studies have contributed to improving control chart pattern recognition (Guh & Tannock 1999; Cheng 1997; Perry, Spoerre, & Velasco 2001; Xanthopoulos & Razzaghi 2014).

A knowledge-based design approach extends traditional rule-based methods for control chart pattern recognition. In this approach, expert systems use if-then rules to automate pattern recognition, reducing reliance on human operators. These systems analyze various pattern characteristics, such as the number of mean crossings, consecutive points on one side of the centerline, and trend slopes. Research efforts by Evans and Lindsay (1988), Duc Truong Pham and Oztemel (1992), and Swift & Mize (1995) focused on applying expert systems to X-bar control charts, demonstrating their potential to improve decision-making in process monitoring. Beyond rule-based methods, pattern-matching algorithms offer another approach for controlling chart pattern recognition. These algorithms compare observed process patterns with predefined abnormal patterns, enhancing automated monitoring capabilities. Studies by Evans & Lindsay (2005), Pham & Oztemel (1994), Swift & Mize (1995), and Guh & Hsieh (1999) explored such techniques, showing their effectiveness in identifying deviations with greater accuracy. As industrial processes become more complex, integrating expert systems and pattern-matching algorithms with control charts can help improve process stability while reducing false alarms and manual interpretation errors. Alghanim (1995) addressed the challenge of detecting process deviations in quality control charts by formulating it as a pattern recognition problem and automating decision-making through statistical and neural pattern recognition techniques. They modeled the control chart as a discrete-time signal composed of random Gaussian noise and unnatural patterns, where the latter indicated an external disturbance requiring intervention.

## Review of Multivariate Control Charts

Multivariate control charts are designed for applications when the process or product quality requires a simultaneous assessment of multiple quality characteristics. For example, a die-casted part may contain multiple locational and dimensional quality characteristics. Compared to applying one control chart for each quality characteristic, a multivariate control chart considering all QCs provides better detection power and prevent false alarms because the relationships among quality characteristics are taken into consideration. The most commonly used multivariate control chart is Hotelling $T^2$ control chart (Hotelling 1947). Equation (1) shows the Hotelling T2 statistics.

$$T^2 = (x - \mu)' \Sigma^{-1} (x - \mu) \qquad (1)$$

where x is the vector of quality characteristics, μ is the mean of x, and ∑ is the variance-covariance matrix of $x$. If a sample n>1 is implemented, the $x$ can be replaced by the vector of the sample averages. When both μ and ∑ are known and the $x$ is multinormal distributed, $T^2$ follows a Chi-square distribution. Then, the upper control limit of the $T^2$ control chart can be set according to the percentile of the Chi-square distribution. When μ and ∑ are unknown, the sample mean and covariance matrix can be estimated. In this case, both F and Beta distributions should be used to compute the control limits.

Similar to its univariate counterpart, Hotelling $T^2$ control chart is not sensitive to small process shifts. Therefore, multivariate control charts such as Multivariate EWMA (MEWMA) and Multivariate CUSUM charts were developed. Crosier (1988) proposed the first multivariate CUSUM, while Lowry et al. (1992) provided guidelines for using multivariate EWMA control charts. Chang and Zhang (2008) proposed a multivariate EWMA control chart for variance shift detection of multivariate autocorrelated processes. Zhang and Chang (2008) extended the multivariate EWMA charts using individual observations for process mean and variance simultaneously. When the dimension of simultaneous quality characteristics is large, the multivariate control charts are difficult to set up since the number of unknown parameters in the variance-covariance matrix is large. It requires a large number of Phase I data sets to estimate all unknown parameters in the variance-covariance matrix. Assuming that all covariance parameters are the same or have some structure, the number of parameters can be dramatically reduced. Another popular solution is the use of the PCA (principal component analysis) technique and its variants ( Wang et al. 2023) for dimension reduction. When the samples are autocorrelated, residuals of the forecasting models are needed for process monitoring. Hajarian et al.(2020) proposed a dynamic PCA method to decrease the effect of autocorrelation and adaptive behavior of moving-window PCA to control non-stationary quality characteristics.

**Review of Fault and Anomaly Detection**

As data collection in complex processes like chemical manufacturing increases, the dimensionality of process parameters has also grown. Ge, Ling, and Chung (2013) defined process monitoring as fault detection and diagnosis, primarily focusing on identifying deviations from standard operating conditions. Traditional fault detection relies on multivariate statistical methods such as principal component analysis (PCA) and partial least squares (PLS), which assume normal data distributions. To relax this assumption, revised methods include independent component analysis (ICA), Gaussian mixture models (GMM), artificial neural networks (ANN), support vector machines (SVM), and support vector data description (SVDD). Nonlinear process monitoring uses nonlinear PCA, kernel PCA, and localized PCA (Kerschen and Golinval 2002).

Anomaly detection, tracing back to the 1960s (Grubbs 1969), identifies deviations from normal operations, including unknown failures. Unlike fault detection, which relies on predefined rules and known failure modes, anomaly detection employs statistical and machine learning techniques to detect novel deviations (Chandola, Banerjee, and Kumar 2009). It is widely used in domains such as manufacturing, fraud detection, cybersecurity, and military surveillance.

Traditional anomaly detection techniques use statistical and rule-based methods like Shewhart, CUSUM, and EWMA control charts, hypothesis testing, and Gaussian-based models (Savage et al. 2014; Qiu 2020). Non-parametric approaches such as Kernel Density Estimation (KDE) (Chen 2017) and distance-based methods like k-Nearest Neighbors (k-NN) and DBSCAN (Çelik, Dadaşer-Çelik, and Dokuz 2011) also detect anomalies based on data proximity. While effective in low-dimensional datasets, these techniques struggle with large-scale and high-dimensional data (Patcha and Park 2007, Munir et al. 2019). Lee et al. (2012) automated anomaly detection in DBMS performance monitoring using SPC charts and differential profiling, reducing manual effort by 90%. Kim and Scott (2012) improved KDE-based anomaly detection through a Robust Kernel Density Estimator (RKDE) with M-estimation, efficiently computed via an iteratively re-weighted least squares (IRWLS) algorithm. Schubert, Zimek, and Kriegel (2014) proposed a generalized density-based outlier detection method that decoupled density estimation from outlier detection, improving adaptability. Angiulli and Pizzuti (2002) enhanced distance-based outlier detection using a Hilbert space-filling curve for efficient search space linearization.

Unlike global methods, another approach detects anomalies or outliers based on local density variation. Various outliers include unusual bank transactions, spiked traffic during a cyberattack, elevated frequency of failures in manufacturing processes, and unusual patient behaviors caused by rare diseases. Breunig et al. (2000) introduced the Local Outlier Factor (LOF) method, which assigned an outlier score based on relative isolation in local neighborhoods, making it more effective than binary classification. Tang et al. (2002) refined LOF with the Connectivity-Based Outlier Factor (COF) method by incorporating connectivity relationships for improved anomaly detection. Papadimitriou et al. (2003) proposed the Local Correlation Integral (LOCI) method, capable of detecting both individual anomalies and micro-clusters while providing detailed local data insights. They further introduced aLOCI, an efficient approximation for fast anomaly detection with near-linear complexity.

**Review of Profile Monitoring**

The SPM methods reviewed so far focus on quality measures that can be quantified by a number (i.e., the univariate cases) or a vector of numbers (i.e., multivariate cases). However, in some cases, the quality characteristic consists of quality measures forming a curve called profile, a function of the quality measure over time or space. For example, Walker and Wright (2002) studied the vertical density profile over the depth of engineered wood boards. Over 314 density observations were taken 0.002 inches apart from a profile. Jin and Shi (1999) used a wavelet method to reduce the data dimension of the stamping force profile. Chang et al. (2012) studied curing process monitoring of temperature profiles in an autoclave for high-pressure hoses.

A large quantity of literature has been published to address two main types of profiles – linear and nonlinear in the last two decades. For linear profiles, simple linear regression models are often used to fit the profile. Kang and Albin (2000) used EWMA and R charts on the residuals. Kim, Mahmoud, and Woodall (2003) suggested three EWMA charts on the intercept, slope, and process standard deviation, respectively. Building on these methods, Arkat, Abbasi, and Niaki (2007) proposed the GLT/R chart to monitor profile coefficients and error variance, offering robustness even when explanatory variables were not fixed. Saghaei, Mehrjoo, and Amiri (2009) proposed a CUSUM-based method specifically designed to monitor simple linear profiles. An extension to simple regression approaches is the use of polynomial regression models known as the response surface models. Ahmadi Yazdi et al.(2024) developed Bayesian multivariate CUSUM control charts for monitoring multivariate linear profiles and demonstrated their superior performance over non-Bayesian counterparts using average run length (ARL) evaluation. Additionally, they proposed an informative prior generation method, leveraging historical Phase I data to improve Phase II monitoring. Shojaee et al. (2024) proposed an integrated model that combines linear profile monitoring using a Hotelling $T^2$ control chart, production cost evaluation, and maintenance policies to minimize total costs. They implemented the particle swarm optimization algorithm (PSO) to determine optimal decision variables.

Control charts are widely used to monitor the residuals of fitted models in quality control. Several recent studies have focused on monitoring linear profiles .Eyvazian et al. (2011) focused on monitoring multivariate multiple linear regression profiles in Phase II SPC, proposing four monitoring methods to account for correlations between response variables. Amiri et al. (2014) extended this by addressing fault diagnosis in multivariate profile monitoring, identifying which profiles and parameters changed during an out-of-control process. Mahmood et al. (2018) introduced new control charts (Max-EWMA-3 and SS-EWMA-3) for simple linear profile monitoring, improving sensitivity in detecting slope, intercept, and error variance shifts. Noorossana et al. (2015)) examined the impact of random explanatory variables on Phase II linear profile monitoring, using ARL criteria to evaluate detection performance. Meanwhile, Chiang et al. (2017) proposed the MEWMA-SLP control chart to handle within-profile autocorrelation in simple linear profiles. Khalafi, Salmasnia, and Maleki (2020)) investigated measurement errors in Phase II monitoring of simple linear profiles, proposing ranked set sampling, multiple measurements, and increased sample size to improve detection accuracy.

Roshanbin, Ershadi, and Niaki (2022) focused on economic-statistical design for simple linear profiles, optimizing cost efficiency and statistical performance using the Lorenzen–Vance cost function and NSGA-II optimization. Khedmati and Niaki (2022) worked on robust parameter estimation in multistage processes, introducing Huber's M-estimator and the MM estimator to mitigate the cascade effect and improve Phase I SPC. Fazelimoghadam, Ershadi, and Niaki (2020) explored economic design in linear profiles, optimizing sampling parameters and control limits for cost-effective monitoring. These studies demonstrate the effectiveness of control charts in tracking deviations from model assumptions and ensuring process stability. Li and Tsai (2023) emphasized the importance of profile monitoring in assessing the stability of functional relationships in manufacturing processes. They introduced the Tweedie exponential dispersion process model to account for within-profile correlation, addressing a key limitation of previous studies. Their simulation results demonstrated that the proposed EWMA-based control charts outperformed existing methods in detecting changes in both linear and nonlinear profiles, with real-world examples validating their effectiveness.

For profiles that are too geometrically complex for simple regression models, several approaches have been proposed. Complex (or nonlinear) profiles examples include temperature-viscosity curves (Garschke et al. 2013), reaction kinetics (Deem et al. 2023), voltage-current characteristics in electronics such as LEDs (Zheng, Chen, and Luo 2018), and fermentation profiles of microorganisms over time (Quinn-Bohmann et al. 2024). Williams, Woodall, and Birch (2007) used $T^2$ charts on the coefficients of the nonlinear regression fits to the successive sets of profile data. Any nonlinear functions can be approximated by a polynomial model with sufficient power of order. Walker and Wright (2002) proposed additive models to compare various non-linear curves. Jin and Shi (1999) and Ding, Zeng, and Zhou (2006) first applied a wavelet method to transform a profile into high- and low-frequency channels. The number of coefficients in these channels was much smaller than the points in the profile of interest. Then principal component (PCA) and independent component (ICA) techniques achieved dimension reduction. Naeem, Ali, and Shah (2025) proposed a Functional Exponentially Weighted Moving Average (FEWMA) control chart for Phase-II monitoring of nonlinear profiles, leveraging cubic B-splines to address high dimensionality issues in classical multivariate control charts. Riaz et al. (2025) proposed new SPC methods using double exponentially weighted moving average (DEWMA) statistics to enhance the monitoring of linear profiles by tracking the intercept, slope, and error variance. Chang and Yadama (2010) further introduced B-splines to fit the smooth curve generated by the wavelet low-frequency channel. After fitting a complex non-linear profile, a sequence of charge points could fully define the trained cubic B-spline. When the number of change points was small, any multivariate SPM methods, such as Hotelling $T^2$ and MEWMA charts, could be directly applied. For the cases where the profiles were very complex (e.g. the tonnage profile from the metal stamping process), a profile could be segmented and residuals were generated from the corresponding B-spline model within each segment. Then, any multivariate control chart could be used to compute the mean square errors based on residuals from multiple segments. Please refer to

Woodall et al.'s (2004) review paper for other methods involving nonparametric regression for profile monitoring.

Chang et al. (2014) proposed a real-time detection method on a wave profile before the entire profile was generated. Since the condensation water temperature collected over time forms the wave profile, early detection of an out-of-shape profile can lead to process fault diagnosis and save costs incurred from unnecessary curing time when a faulty curve is spotted in the early stages. Chou, Chang, and Tsai (2014) addressed the situation where multiple profiles needed to be monitored simultaneously. Other methods can be found in Maleki, Amiri, and Castagliola.'s (2018) survey paper.

For applications where the quality measures may be binary or counts over time, non-linear profile methods have been studied. These kinds of profiles include the following examples. The airline industry is interested in monitoring the number of incidents as the proportion of total flights over time (Yeganeh et al. 2024). A car manufacturer is interested in learning the patterns of monthly warranty claims since launch (Chenglong Li et al. 2020). Usually, a Poisson distribution is suitable for the quality characteristics. Mohammadzadeh, Yeganeh, and Shadman (2021) proposed a generalized likelihood ratio and support vector regression methods for logistic profiles of dose responses. Shang, Wang, and Zhang (2018) used nonparametric regressions using generalized likelihood ratios for attribute responses such as binary or Poisson data. Ghasemi, Hamadani, and Yazdi (2025) proposed three new methods based on the multivariate homogeneously weighted moving average (MHWMA) control chart to enhance the monitoring of multivariate simple linear profiles. Other works in this area include Chen and Nembhard (2011) for high-dimensional (HD) control chart based on the adaptive Neyman test statistic to monitor both linear and nonlinear profiles, particularly in the presence of stationary noise, Koosha and Amiri (2011) for autocorrelated logistic regression profiles, Amiri et al. (2012) for parameters reduction method to monitoring multiple linear regression profiles, Derakhshani et.al. (2020) for Poisson regression profiles in multi-stage processes, and Isazadeh Khanghah et al. (2023) for gamma regression profiles.

Zhang et al. (2018) highlighted the challenge of real-time process monitoring in advanced manufacturing systems with high-dimensional sensor data affected by noise and cluster-wise correlations. To address this, they proposed a multiple profiles sensor-based monitoring system that preprocesses signals, extracts features using multichannel functional principal component analysis (MFPCA), and applies a top-R strategy for scalable fault detection. The proposed framework was validated using real manufacturing system data, demonstrating its effectiveness in anomaly detection. Hadidoust, Samimi, and Shahriari (2015) used the spline smoothing method for modeling and apply the Hotelling $T^2$ control chart to monitor process variations. Their simulation studies demonstrated that the proposed method effectively detected large global shifts and was highly sensitive to local shifts, making it a robust approach for non-linear profile monitoring. Mortezanejad et al. (2024) proposed two methods—linear regression and the maximum entropy principle—for monitoring the intercept and slope of a simple linear profile instead of just the process mean. Their study, using Hotelling $T^2$ statistic and simulation comparisons, showed that while both methods perform similarly, maximum entropy was more effective in detecting changes in pharmaceutical processes

**Review on Multistage QE Applications**

The SPC literature reviewed so far applies to univariate, multivariate, or profile quality characteristics in a single stage. The assumption is that previous operations leading up to the stage under monitoring are independent of the current stage or not considered in current process monitoring. However, many manufacturing or service operations violate this assumption. For example, in 3D printing and semiconductor manufacturing, minor faults in early operation stages may not be detected immediately but may affect product quality in subsequent stages (Bisheh, Chang, and Lei 2021). Tsung, Li, and Jin (2006) provided a comprehensive review of SPC for multistage manufacturing and service operations. The main challenge of applying SPC for multistage operations lied in the process modeling.

The most adopted statistical modeling methods include linear regression models (Basseville and Nikiforov 1993) and linear state-space models . Assuming a process has a total of N stages, the quality measures $y_i$ at stage *i* can be modeled as the following two-level equations (equation (2) & equation (3)).

$$x_i = A_{i-1}x_{i-1} + v_i \qquad (2)$$
$$y_i = C_i x_i + w_i \qquad \text{for } i = 1,2,…,N \qquad (3)$$

where $x_i$ is the unobservable product or process quality measures, $y_i$ is the quality measures, $v_i$ is the measurement error, and $w_i$ is the common-cause variation and unmodeled errors. Both $v_i$ and $w_i$ are assumed Normally distributed. $A_{i-1}$ and $C_i$ are assumed known constant matrices and require process expert knowledge. The main objective of the SPM approach based on this state-space model is to detect if the means of $y_i$ at *i* = 1,2,…,N have changed from the known levels. Note that this two-state model is often used in signal processing known as the Kalman Filter model (Kalman 1960).

Although this state-space model provides a general framework for SPM of multistage processes, it may not be easily adapted to processes when the number of stages is large. Zhou, Huang, and Shi (2003) described an engine-head operations that involve more than 30 stages. In this case, it was difficult to model all stages in sufficient detail for fault detection or isolation. The design of a multistage process monitoring and diagnosis scheme consisted of two key considerations: what process variables or quality characteristics needed to be monitored, and where monitoring should take place. If the single-stage SPM method such as EWMA or CUSUM was applied to each $y_i$, the simultaneous use of all charts may lead to unsatisfactory Type I or Type II errors depending on how the control limits of each individual chart are set up. Overcoming this drawback, Xiang and Tsung (2008) proposed using a group of EWMA charts for the one-step forecast errors of the model of each stage. In addition, the EM algorithm was applied to estimate model parameters for practical implementation of the proposed state-space model. Please refer to Shang, Tsung, and Zou (2013) for applications where the process outcomes are binary.

Yeganeh et al. (2024) emphasized the importance of monitoring healthcare processes to detect changes and unnatural conditions early, ensuring patient safety. They proposed a Multistage Process Monitoring (MPM) control chart tailored for healthcare data, integrating machine learning with statistical control charts to enhance detection capability (Intelligent Control Charting, ICC). Their Monte Carlo simulations and real-world application to thyroid cancer surgery confirmed the effectiveness of the proposed approach in healthcare process monitoring. Jalilibal et al. (2024) stressed the significance of overseeing multi-stage processes with high-dimensional data streams across manufacturing and non-manufacturing sectors. In their review, they underscored the necessity of cause-selecting control charting methods for precisely identifying process disruptions.

Bahrami, Niaki, and Khedmati (2021) studied multivariate profile monitoring in multistage processes, where quality characteristics evolve across multiple stages of production. They proposed a general model using the multivariate U transformation approach to eliminate the cascade effect between stages, improving monitoring accuracy. Three control schemes were developed for Phase II monitoring of multivariate simple linear profiles, with performance evaluated through average run length (ARL) simulations. Also In a review paper, Jiang, Yan, and Huang (2019) emphasized data-driven distributed monitoring methods for industrial plant-wide processes, emphasizing the Multivariate Statistical Process Monitoring (MSPM) approach for maintaining stable operations. They introduced the Data-driven Multivariate Statistical Plant-Wide Process Monitoring (DMSPPM) framework, which decomposes complex processes into subprocesses for local monitoring using PCA, CCA, and a Bayesian-based method for joint analysis. They also discussed process variable decomposition methods and key challenges, such as complex process characteristics, imbalanced data, multisource signal fusion, and fault propagation analysis. The study concluded that further research was needed to enhance theoretical and practical applications of distributed monitoring in large-scale industrial settings.

**Review of Image-Related QE Applications**

Machine vision methods have consistently improved image-based analysis, particularly in areas such as defect detection and surface assessment. These methods leverage traditional image processing techniques to ensure reliable and efficient quality control across various industries. For instance, computer vision technologies offer a non-contact approach for evaluating surface roughness and other image characteristics, enabling real-time monitoring and inspection in manufacturing processes (Chang and Ravathur 2005). Specifically, they used wavelet transformation on machining surfaces and established the relationship between images and surface roughness measure.

Duchesne, Liu, and MacGregor (2012) proposed a partial least squares regression (PLSR) to analyze complex multivariate images. The combination of MIA and image PLS, implemented in LabVIEW and MATLAB, demonstrates how data can be decomposed differently than traditional image PCA, which is vital for technical image analysis in various industries (Lied, Geladi, and Esbensen 2000). Examples of MIA applications include medical imaging such as MRI and CT scans. These techniques illustrate how traditional, non-AI methods, such as feature extraction and image segmentation, play a key role in quality control despite the increasing complexity of high-dimensional data (Montalbán, Juan, and Ferrer 2011).

Techniques such as feature extraction and Principal Component Analysis (PCA) are used to reduce the dimensionality of image data and focus on the most significant features, making it easier to detect defects and monitor processes (Adegoke et al. 2022). Optical inspection and pattern matching, widely used in manufacturing, further automate defect identification based on predefined rules. These traditional methods, such as Fourier Transform Analysis and thresholding, remain effective in detecting periodic patterns or noise in images. However, they may lack the flexibility required for more dynamic environments (Sahoo et al. 2015; J. Lyu and Chen 2009; Seim, Andersen, and Sandberg 2006). Despite the rise of AI-based approaches in quality control applications, non-AI methods remain cost-effective and widely implemented, especially in industries dealing with straightforward processes or lower-dimensional data, such as healthcare and electronics (Zhou, Sun, and Shi 2006; Colosimo, Semeraro, and Pacella 2008). Together, these methods highlight the importance of non-AI techniques in maintaining consistent image quality and enhancing process monitoring.

# A Review of AI-Enabled Methods for Process Monitoring and Anomaly Detection

**A Taxonomy of AI-based and Statistics-based Methods Addressing SPC Applications**

**Fig. 1** presents an overview of AI methods (AI-SPC) and traditional statistical approaches (T-SPC) for statistical process monitoring applications. The term SPC instead of SPM reflects the possibility that AI methods may achieve true process control by adjusting process parameters to restore a process to an in-control state. The taxonomy of AI-SPC is further divided into four categories: classification, pattern recognition, time series application, and generative AI. The ending blocks represent various ML or AI techniques for SPC. T-SPC has two categories in variable and attribute monitoring. The ending blocks are various control charts. Details of the crucial AI-SPC techniques will be discussed in the following sections. Note that the classification of AI-SPC methods primarily focuses on ML/AI techniques. In contrast, the classification of T-SPC focuses on quality characteristics such as types of quality characteristics (variables or attributes), dimensions (univariate, multivariate, or profiles), and locations (local or multiple stages). The following sections will review AI-SPC methods under the T-SPC categories.

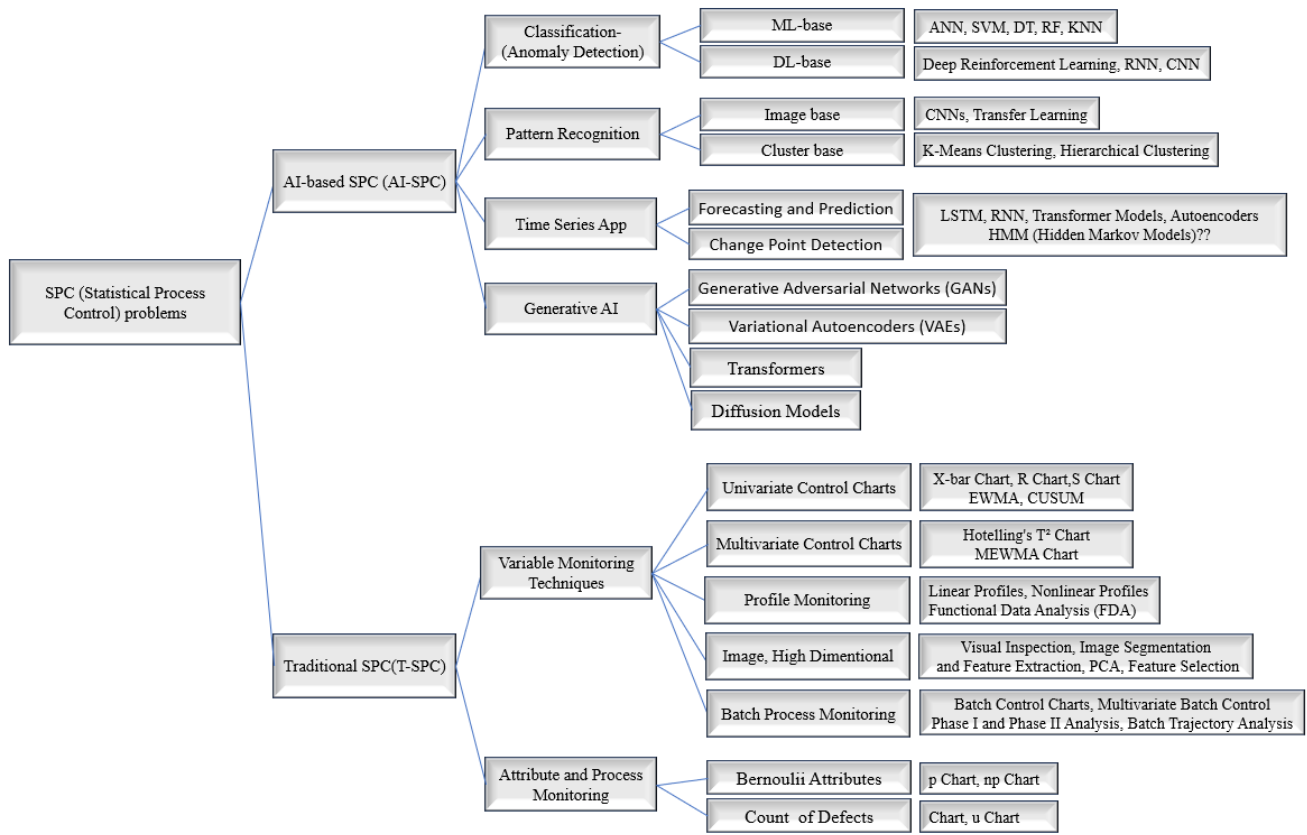

Fig. 1. A Taxonomy of AI-based and Statistics-based Methods Addressing SPC Applications

**Early Neural Network Control Charts for Univariate Quality Characteristics**

Many attempts have been made to apply neural networks to process control since the 1990s [Antognetti 1991; Schuster and Schuster 1992]. These neural network process control tools provided an alternative way to monitor process variability, which was traditionally statistically based. Ho and Chang (1999) developed a combined fuzzy set and neural network scheme to monitor a process and estimate mean shift or variance change magnitudes after an out-of-control situation occurred. Simulation results (Chang and Aw 1994; Chang and Ho 1999; Ho and Chang 1999) showed that the proposed neural network control charts, called C-NN charts, perform well compared to other traditional statistical-based control charts regarding average run length (ARL) and correct classification rates. In addition, C-NN charts were robust to assumption violations, such as autocorrelation or non-normal data. These early ANN-based SPC applications were mainly designed for univariate quality characteristics. Although these ANN control charts outperformed statistics-based control charts such as X-Bar, EWMA, or CUSUM in some situations, the adoption has not been widespread. One of the main reasons is the need for a large number of samples for ANN training. Traditional control controls usually require approximately 100-125 observations to set the control limits, while an ANN chart may require a lot more. We used 4000 simulated observations from a standard Normal distribution to train an ANN chart comparable to an X-bar chart. The architecture of the ANN was 10-20-20-1. However, the number of training observations could be dramatically reduced when the AI transfer-learning method was adopted. The main idea of transfer learning was a foundation NN model trained on large samples, in this case, Normally distributed samples. Then, additional layers were added to the foundation NN and trained from a much smaller sample only on additional layers.

**AI-Enabled SPM for Multivariate Applications**

Bersimis, Sgora, and Psarakis (2017) compared Neuro-Fuzzy Inference (ANFIS), Support Vector Machine (SVM), and Neural Networks (ANN) control charts to various multivariate control charting methods, including the Doganaksoy, Faltin, and Tucker (DFT) algorithm, the Mason, Young, and Tracy's (MYT), and Murphy out-of-control algorithm. These control charts aimed to identify which dimensions contributed to the out-of-control process. Three types of variance-covariance matrices were considered in their simulation studies of three and five variables. The ANN architecture included three layers with 8 or 12 nodes in the input layer, 12 in the hidden layers, and 3 or 5 in the output layers (e.g., the smallest architecture is 8-12-3). The Sigmoid function was adopted for the activation function for both hidden and output layers. The chart performance regarding successful identification percentages showed that the ANN approach outperformed in some cases but lags in the other methods. For example, the ANN could detect a one-variable shift of -3 sigma with a 91.4% rate but only detect 7.8% when the shift was +3 sigma. The MYT algorithm, on the other hand, was capable of 77.1% and 76.1% for -3 and +3 sigma shift cases, respectively. However, when the process shift was very small, such as 0.5 sigma, the successful identification percentages were 45.0%, 17.4%, and 13.9% for ANN, MYT, and DFT, respectively. Yeganeh et al. (2023) implemented these machine learning-based control charts on the network surveillance problem where the numbers of communications between nodes in the network were under surveillance.
    Salehi, Bahreininejad, and Nakhai (2011) proposed a learning-based framework for multivariate manufacturing processes. The proposed framework consisted of two modules. Specifically, a support vector machine classifier detected unnatural process patterns in the first module. Then, three neural networks identified potential out-of-control patterns in mean shifts, trends, or cycles. Note that each neural network served as an AI-agent that focused on one specific task.

**AI-Enabled SPM for Profile Monitoring Applications**

The integration of Artificial Intelligence (AI) and Machine Learning (ML) into Statistical Process Monitoring (SPM) has revolutionized profile monitoring by improving the detection of process shifts in linear and nonlinear profiles. Researchers have developed advanced adaptive control charts, nonparametric regression techniques, and AI-based models to enhance monitoring capabilities. Traditional statistical methods, such as EWMA and Hotelling $T^2$ charts, are now being complemented with AI-driven approaches to handle complex and non-normally distributed data.

Several studies have explored ANNs and machine learning models for profile monitoring. Sabahno and Amiri (2023) applied ANN, SVR, and RFR to monitor generalized linear model (GLM) profiles, incorporating an adaptive scheme that dynamically adjusted sample sizes and intervals. Their findings suggested that ANN-based monitoring performs comparably to SVR and RFR, with the potential for further improvement through deeper architectures. Similarly, Awad, AlHamaydeh, and Faris (2018) utilized ANNs combined with Hotelling's $T^2$ charts to model variable relationships, demonstrating superior fault detection in reinforced concrete monitoring systems. Additionally, Hosseinifard and Abdollahian (2010) proposed a supervised feedforward neural network to detect and classify drift shifts in linear profiles, showing substantial average run length (ARL) performance.

Nonparametric approaches have also been widely adopted for profile monitoring. Hric and Sabahno (2024) introduced an ML-based framework for monitoring GLM profiles with Binomial, Poisson, and Gamma distributions, showing that control chart performance varies based on response distribution and training method. Timme et al. (2024) developed a nonparametric method using regression tree ensembles and random forests, incorporating the Kolmogorov-Smirnov statistic for detecting change points in multivariate profiles. Additionally, Pan and Lu (2019) addressed nonlinear, non-normally distributed profiles by proposing SVR-based monitoring with a revised spatial rank EWMA (RSREWMA) chart, demonstrating its effectiveness in detecting process shifts when common fixed design (CFD) was not feasible. Abdel-Salam, Birch, and Jensen (2013), addressed model misspecification in linear and nonlinear profiles. Their Mixed Model Robust Profile Monitoring (MMRPM) method integrates parametric (P) and nonparametric (NP) approaches.

Several studies have focused on profile monitoring for non-normal data. Li, Pan, and Liao (2019) emphasized the necessity of nonparametric regression models in modern manufacturing due to nonlinear relationships in profile data. They introduced an SVR-based EWMA control chart that significantly improved shift detection. Yeganeh et al. (2024) tackled Poisson profile monitoring in Phase II, proposing an SVR-based control chart with evolutionary training that outperformed traditional methods such as LRT and MEWMA. These advancements highlighted the growing need for AI and ML-driven techniques in profile monitoring, offering more adaptive, data-driven, and robust solutions to quality control challenges.

### AI/ML Methods for Pattern Recognitions of Control Chart Patterns

Hachicha and Ghorbel (2012) analyzed 120 studies from 1991 to 2010. They proposed a conceptual classification scheme to organize Control Chart Pattern Recognition (CCPR) research and highlight key trends. Most studies focus on IID process data, while recent work explores multivariate process shifts using AI techniques such as ANNs, wavelet-denoising, and hybrid models with decision trees and particle swarm optimization. Performance evaluation in CCPR relied mainly on recognition-accuracy metrics, as ARL-based criteria are inadequate for concurrent pattern detection. The study emphasized the growing need for hybrid AI models, improved concurrent pattern recognition, and better evaluation methods to advance CCPR research. Garcia et al. (2022) searched the Web of Science and Scopus databases to identify control chart patterns. Out of the 44 papers identified, they found that the most popular method used was ANN at 30% followed by SVM at 23% and DNN at 11%. Almost 78% of the work was based on normally distributed data. However, the control chart pattern recognition problem has not garnered the attention of the SPC research community. Less than five papers were published each year between 1999 and 2022. No papers were published between 2000 and 2003.

Cheng (1997) proposed an ANN framework for control chart pattern analysis. Two ANNs work in tandem to first detect unnatural patterns and then to classify patterns according to the underlying causes. The proposed framework works well when the signal-to-noise ratio is low. Gauri and Chakraborty (2008) proposed the use of ANN for control chart patterns. Most classical methods for control chart pattern recognition were heuristics-based, but consistency was often the main concern. On the other hand, the proposed ANN approach could be used to extract distinct features for the patterns to be identified. The authors reported better performance concerning both recognition accuracy and consistency. The control chart patterns considered included stratification, systemic, cyclic, upward shift, downward shift, increasing trend, and decreasing trend. Zaidi et al. (2023) analyzed abnormal patterns in a $T^2$ control chart using ANNs. Cheng et al. (2023) developed a deep learning-based classification model to recognize control chart patterns in multivariate processes using a multi-channel deep convolutional neural network (MCDCNN). Their model integrated both 1D and 2D representations of control chart data, significantly improving classification accuracy, particularly for detecting non-random patterns. The model outperformed traditional techniques by 10%, offering notable improvements in intelligent SPC, especially in complex, multivariate scenarios.

Addeh, Khormali, and Golilarz (2018) proposed an optimized radial basis function neural network (RBFNN) with association rules for feature selection and the bees algorithm for improved learning, extending Control Chart Pattern (CCP) recognition to eight patterns. Similarly, Ebrahimzadeh and Ranaee (2010) leveraged wavelet packet entropy for feature extraction, enhancing neural network classifiers using particle swarm optimization (PSO). Guh and Shiue (2010) introduced a time delay neural network (TDNN) that improved the recognition of dynamic patterns over traditional backpropagation networks, while Masood and Hassan (2010) explored enhancements in ANN-based recognition through feature selection and wavelet denoising techniques.

SVM-based approaches have also been widely utilized for CCPR. Khormali and Addeh (2016) developed a multiclass SVM model integrated with type-2 fuzzy c-means clustering (T2FCM) and optimized using the cuckoo optimization algorithm (COA), improving unnatural pattern recognition. Zhou, Jiang, and Wang (2018) introduced a fuzzy SVM with a hybrid Gaussian-Polynomial kernel, optimized with genetic algorithms (GA), outperforming traditional ML models. Chowdhury and Janan (2020) compared SVM with statistical correlation-based methods, highlighting that while SVM achieved higher accuracy, it was computationally intensive. Xanthopoulos and Razzaghi (2014) addressed imbalanced datasets in CCPR by introducing Weighted Support Vector Machines (WSVM), which prioritized minority class samples, improving fault detection accuracy over traditional SVM-based K charts.

Recent advancements have focused on deep learning models for CCP recognition. Derakhshi and Razzaghi (2024) introduced a cost-sensitive bi-directional LSTM model with an adaptive weighting strategy and bi-objective early stopping, improving process monitoring in biomanufacturing and wafer production. Xue et al. (2023) developed a multi-feature fusion CNN (MFF-CNN) incorporating SMOTE for data balancing, outperforming traditional classifiers in recognizing abnormal CCPs. Chiu and Fu (2024) proposed a CNN-LSTM hybrid model for online anomaly detection in SPC-EPC environments, achieving 99.83% accuracy, while Yu and Zhang (2021) introduced an optimized CNN-LSTM network with a genetic algorithm-based hyperparameter tuning method, enhancing real-time fault detection in intelligent manufacturing.

Beyond deep learning, other AI-driven methods have also shown promise in CCPR. Lee et al. (2022) introduced a spectral clustering-based SVM for CCP recognition under gamma-distributed datasets, addressing the limitations of traditional methods under non-normal distributions. Haghtalab,

Xanthopoulos, and Madani (2015) developed an unsupervised consensus clustering framework, achieving over 90% accuracy in anomaly detection. Alwan et al. (2023) proposed an ensemble-based CCPR method, combining decision trees, ANN, linear SVM, Gaussian SVM, and k-NN to detect small process variations. Pelegrina, Duarte, and Jutten (2016) explored Blind Source Separation (BSS) techniques for recognizing overlapping CCPs, demonstrating the importance of efficient feature extraction. Overall, AI and ML have revolutionized CCP recognition by enhancing fault detection, improving process monitoring, and enabling real-time decision-making, with future research focusing on digital twins, handling imbalanced datasets, and optimizing AI models for industrial-scale applications.

**AI-enabled Image-Related QE Applications**

The review of image-related Quality Engineering (QE) applications underscores the transformative role of Artificial Intelligence (AI), with particular emphasis on Convolutional Neural Networks (CNNs) and their impact on modern quality control and monitoring practices. CNNs have emerged as the foundational technology for automating visual inspection processes, enabling real-time, high-precision analysis of visual data to detect and diagnose defects within complex manufacturing workflows (Ding et al. 2024). These deep-learning models have proven exceptionally effective in identifying subtle and intricate visual irregularities that may elude human inspectors and are often undetectable by traditional statistical methodologies (Zhen, Paynabar, and Shi 2023). By leveraging the hierarchical structure of CNNs, which captures both low-level and high-level features in image data, these models are adept at recognizing patterns, shapes, and textures that signify potential defects, thereby enhancing the robustness and accuracy of quality monitoring systems (Duan and Mou 2021)

In the domain of medical imaging, Ghafariasl, Zeinalnezhad, and Chang (2025) proposed a neural network framework that integrates transfer learning and deep learning to extract and diversify features from mammographic images. The framework combined three pre-trained networks: ResNet-50, VGG-19, and ResNet-152V2, with additional layers fine-tuned using both segmented and non-segmented images. The ensemble of these networks achieved an AUC of 0.86 for segmented images and 0.80 for non-segmented images. Furthermore, the stacked generalization model that combined features from all three networks achieved an AUC of 0.89, showing significant potential for improving accuracy in medical image classification tasks. Similarly, Lai et al. (2023)'s approach in Skin Cancer Diagnosis (SCD) employs machine learning to identify abnormal patterns, akin to how Statistical Process Control (SPC) monitors process deviations; both methods focus on optimization and classification to improve accuracy in detecting critical anomalies.

AI-powered image-related Quality Engineering (QE) applications leverage deep learning models like LSTM and Bi-LSTM to enhance predictive accuracy and operational efficiency in modern industrial processes. Note that the control-chart patterns reviewed in the previous section can also be considered a special case of image-based QE applications. In the context of SPC, Zan et al. (2019) introduced an intelligent SPC framework using Bi-LSTM networks for feature learning, significantly improving Histogram Pattern Recognition (HPR) and CCPR. Their approach addressed the limitations of traditional methods by enabling more accurate and efficient detection of abnormal patterns, thus supporting real-time decision-making and automation essential for Industry 4.0. This advancement demonstrated the potential of deep learning to transform quality management in manufacturing environments.

Building on this progress, Aydemir and Paynabar (2020) proposed advanced deep-learning techniques for estimating the time-to-failure (TTF) of industrial systems using degradation images. Their methodology utilized LSTM networks to capture temporal dependencies in image data, offering two innovative approaches: one combines convolutional layers for feature extraction with LSTM layers for TTF prediction, while the other reduces image dimensionality through an autoencoder, followed by LSTM-based TTF estimation. These techniques enhanced predictive maintenance by providing more accurate failure predictions, minimizing downtime, and improving reliability in industrial settings. Together, these studies highlighted the transformative role of deep learning in advancing quality engineering and predictive maintenance in complex industrial systems.

Integrating image-based data with other critical process parameters has catalyzed the development of sophisticated fault detection frameworks, which offer a more comprehensive and resilient approach to monitoring and maintaining quality standards (Chen et al. 2022). In particular, the synergy between image-based methods and process data analytics has led to the creation of multi-modal monitoring systems, where visual data is combined with time-series or sensor data to provide a holistic view of the production process. This multi-faceted approach not only improves defect detection but also facilitates predictive maintenance and real-time decision-making, thereby reducing downtime and improving overall process efficiency (Tao, Leu, and Yin 2020).

Méndez et al (2024) introduced the Novel Intelligent Inspection Method (NIIM) to enhance contour profile quality assessment in CNC milling processes. NIIM integrated a calibration piece, the RAM-Starlite™ vision system, and machine learning techniques to detect and classify form deviations with high accuracy. Their experimental study, using 356 images from 60 machined calibration pieces, demonstrated 96.99% accuracy and low computational requirements, providing valuable insights for improving machining quality.

Furthermore, the continual advancements in deep learning paradigms have significantly expanded the capabilities of these image-centric monitoring systems, particularly in discerning complex patterns and anomalies within high-dimensional datasets. For instance, advanced techniques such as Generative Adversarial Networks (GANs) have been employed to generate synthetic data that augments training datasets (Kornish, Ezekiel, and Cornacchia 2018), thereby improving model robustness and generalization to unseen defects. Details of GAN will be discussed in the following section.

Similarly, applying Transfer Learning (TL) for deep learning model building has revolutionized the deployment of CNNs in quality engineering (Pan and Yang 2010). By utilizing pre-trained models on large-scale datasets and fine-tuning them on domain-specific data, TL reduces the need for extensive labeled datasets, which are often difficult and costly to obtain in industrial settings. This approach accelerates the development and deployment of AI-driven QE systems and enhances their adaptability to different manufacturing environments and defect types (Birkstedt et al. 2023).

In addition, other deep learning methodologies such as Recurrent Neural Networks (RNNs) and Long Short-Term Memory (LSTM) networks have been integrated with CNNs to capture temporal dependencies in image data streams, further enhancing the ability to monitor dynamic processes and detect temporal anomalies. This is particularly relevant in applications such as additive manufacturing, where the process evolves, and continuous monitoring is critical to ensure product quality (Ghimire et al. 2021). Bazi et al. (2022) demonstrated how RNNs and CNNs collaborated effectively in Quality Engineering (QE) applications, particularly in tool condition monitoring and wear prediction in manufacturing processes. Their work exemplified the power of the combined CNN-BiLSTM model, which integrated these two architectures seamlessly. Specifically, the CNN component acted as a feature extractor, identifying critical spatial patterns from the data, while the BiLSTM component functioned as a sequence modeler, capturing temporal dependencies inherent in the signals. Together, this hybrid architecture leverages spatial and temporal information to ensure robust and accurate predictions of tool wear, making it a powerful solution for enhancing manufacturing efficiency and precision.

**Generative Adversarial Networks for QE Image Applications**
Generative adversarial networks (GANs) are a class of AI algorithms used in unsupervised machine learning (Goodfellow et al. 2014). GANs involve two neural networks – a Generator and a Discriminator, competing against each other. Specifically, the Generator creates data instances such as an image with random noise. Its goal is to deceive the Discriminator by generating data so realistically that it cannot be distinguished from real data. On the other hand, Discriminator serves as a critic to evaluate data from both the Generator and the real-world dataset. It aims to improve its ability to detect fakes over time. Once trained, the Generator learns to produce more plausible fakes to trick the Discriminator, where the Discriminator improves at spotting fakes. GANs have been used to produce realistic images (Gragnaniello et al. 2021) and to mimic the composing styles of famous composers (Liu 2023).

GANs are powerful imaging application tools capable of generating high-quality, realistic images. They are commonly used in medical imaging for data augmentation, synthesizing diverse datasets to improve model training and generalization. Synthetic data generation alleviates the burden of collecting large sample data for AI model training. For example, the GANs framework has been implemented for surface defect detection. The main challenge of training an ANN for defect classification is the lack of training data for each defect types. The GAN framework provides a mechanism to generate synthesized images. Zhu et al. (2022) used a semi-supervised multitask GAN to identify surface detectors. Multiple GANs were used, and each was responsible for one type of defect. The outcome of the GAN model was either "real" or "fake." Results from multiple GANs were then fed into a classifier to determine the defect type. Du, Gao, and Li (2023) proposed a contrastive GAN with data augmentation for surface defect identification. The proposed method had two components: defect generation and defect recognition. The former produced synthesized defect images while the latter recognized defects under limited data. The proposed framework showed an accuracy rate of 91.84% for a printed circuit board dataset. However, the training of the proposed GAN model was done offline, and the training process was slow. Further computational improvement is needed for online, real-time applications. GANs have also been implemented in various image-based surface defect identification, such as semiconductor wafers (Kim et al. 2021) and steel strips (Ran et al. 2024). GANs also excel in image reconstruction, enhancing low-resolution images or filling missing regions, which is critical in diagnostics and visualization. Furthermore, they enable style transfer, such as converting between imaging modalities (e.g., MRI to CT), ensuring consistent analysis across diverse datasets (Kazeminia et al. 2020).

**AI-Based Anomaly Detection Techniques**
AI-based anomaly detection techniques leverage machine learning and deep learning to dynamically identify anomalies, often outperforming traditional methods in complex scenarios. Supervised models like Support Vector Machines (SVMs) and Random Forests require labeled data, while unsupervised techniques such as Autoencoders, Isolation Forests, and Self-Organizing Maps (SOM) detect anomalies without prior labels (N. Li et al. 2021). Advanced deep learning approaches, including Variational Autoencoders (VAEs), Long Short-Term Memory Networks (LSTMs), and Generative Adversarial Networks (GANs), excel in handling high-dimensional and sequential data (Xia et al. 2022). These AI-based methods are highly adaptive and effective in applications like cybersecurity, healthcare, and predictive maintenance but require careful tuning and substantial computational resources (Xu, Liu, and Yao 2019, Pang et al. 2021, Wang et al. 2020). Anomaly detection in high-dimensional data presents unique challenges, often called the "curse of dimensionality." Thudumu et al. (2020) addressed these issues through a triangular model incorporating detection techniques, data challenges, and big data frameworks. Their study reviewed modern strategies to overcome the limitations of traditional methods and enhance anomaly detection accuracy. Similarly, Koren, Koren, and Peretz (2023) proposed a feature-wise anomaly detection method that independently applied multiple techniques to each feature. Their Noise Ratio (NR) metric improved outlier detection accuracy while avoiding imputation or removal of data points.

Several studies focus on anomaly detection in industrial and manufacturing settings. Lang et al. (2022) introduced a one-class anomaly detection method for semiconductor fabrication, using Kernel Density Estimation (KDE) to classify sensor data anomalies. Their adaptive approach outperformed traditional SPC and OC-SVM models in detecting faults. Mukhtiar, Zaman, and Butt (2024) proposed machine learning-based control charts (MLCCs) to enhance process monitoring, addressing the limitations of traditional control charts by incorporating ML techniques to detect process shifts more effectively. Basora, Olive, and Dubot (2019) review unsupervised methods for time-series anomaly detection in the aviation sector, highlighting neural networks and temporal logic for operational safety improvements. Sheridan et al. (2020) enhanced flight safety analysis using hierarchical clustering and DBSCAN, demonstrating that analyzing the final seconds of the flight approach improved risk identification. West, Schlegl, and Deuse (2023) applied unsupervised k-Means clustering with Dynamic Time Warping for anomaly detection in automotive manufacturing, reducing the need for extensive preprocessing. Machine learning techniques, such as clustering and time-series analysis, and advanced generative models such as Autoencoders, Variational Autoencoders, and GANs, offer enhanced capabilities for detecting anomalies in modern industrial systems (Tran et al. 2022).

Generative models play a crucial role in anomaly detection. Munir, Siddiqui, et al. (2019) introduced DeepAnT, a CNN-based time-series predictor that detected point and contextual anomalies without labeled data, proving effective for IoT applications. Raza et al. (2023) proposed AnoFed, a federated framework integrating transformer-based Autoencoders, VAEs, and SVDD for ECG anomaly detection in digital healthcare. Makhlouk (2018) focused on clustering-based methods in semiconductor manufacturing, integrating decomposition-based algorithms like STL and SAX to enhance process quality control. Hybrid models combining statistical and machine learning techniques improved anomaly detection performance. Zhao et al. (2022) proposed the GT-MSPC model, integrating multivariate statistical process control (MSPC) with TOPSIS and a grey model to enhance fault detection in blast furnace ironmaking. Zheng, Li, and Zhao (2016) combined hypothesis testing with fuzzy set-based techniques to improve early anomaly detection and reduce false alarms. Tayalati et al. (2024) integrated SPC with an LSTM-based Autoencoder for injection molding process monitoring, achieving high accuracy in detecting complex anomalies.

For video anomaly detection, Yang, Liu, and Wu (2021) proposed BR-GAN, a Bidirectional Retrospective GAN that improves prediction-based methods by enhancing spatial and temporal consistency. Chatterjee and Ahmed (2022) reviewed IoT-based anomaly detection challenges, including sensor diversity, data drift, and the absence of ground truth labels. Rosenberger et al. (2022) introduced EEM-KDE, an enhanced KDE algorithm for detecting various anomaly types in edge computing, demonstrating efficiency in low-latency industrial applications. Senthilraja et al. (2024) proposed a dynamic behavioral profiling and anomaly detection framework for Software-Defined IoT Networks (SD-IoT) to enhance security. By leveraging Software-Defined Networking (SDN), the framework continuously monitored IoT device behavior, using machine learning-based anomaly detection to identify security threats. Yan, Paynabar, and Shi (2017) developed a smooth-sparse decomposition method for detecting anomalies in noisy manufacturing images, outperforming traditional techniques in real-time inspection. Qorbani et al. (2025) introduced a method for anomaly detection in photovoltaic (PV) plants using wavelet transformation and robust regression, improving reliability and efficiency. Nguyen et al. (2021) combined an LSTM Autoencoder with a one-class SVM for detecting anomalies in sales data, showing superior performance in multivariate time-series forecasting. Kaplan and Alptekin (2020) explored supervised, unsupervised, and semi-supervised approaches for anomaly detection, applying BiGAN (Bidirectional Generative Adversarial Network) to the KDDCUP99 dataset. Their study introduced training strategies to reduce generator-discriminator dependency, enhancing anomaly detection effectiveness. Finally, Munir, Siddiqui, et al. (2019) proposed FuseAD, a hybrid statistical and deep learning model integrating ARIMA and CNNs for real-time sensor anomaly detection, minimizing system downtime.

Recent advancements in fault diagnosis and anomaly detection have leveraged explainable AI, generative adversarial networks (GANs), and physics-informed machine learning (PIML) to address challenges in industrial AI applications. Brito et al. (2023) introduced Fault Diagnosis using eXplainable AI (FaultD-XAI), an approach utilizing transfer learning from synthetic vibration data and Gradient-weighted Class Activation Mapping (Grad-CAM) with a 1D CNN, enhancing model interpretability for rotating machinery fault diagnosis. Similarly, Tong et al. (2022) tackled small and imbalanced (S & I) fault samples in rolling bearing diagnosis by developing ACGAN-SN, a GAN-based model with spectral normalization, which stabilized training and improves classification accuracy. Hao, Du, and Liang (2022) proposed MFGAN, a multi-resolution fusion GAN that maps normal data into fault space and employs a multi-scale discriminator, effectively augmenting imbalanced fault datasets for industrial bearings.

Beyond GAN-based augmentation, hybrid AI models have emerged to integrate physical laws with data-driven learning for enhanced reliability. Wu, Sicard, and Gadsden (2024) emphasized Physics-Informed Machine Learning (PIML) as a key methodology, embedding domain knowledge into ML models to improve fault detection and predictive maintenance. Their review highlighted how PIML bridged the gap between traditional physics-based models and statistical ML methods, enhancing both accuracy and interpretability. In the realm of networked sensor systems, Li et al. (2019) proposed MAD-GAN, an unsupervised GAN-based anomaly detection model that captured spatial-temporal dependencies among sensors, demonstrating strong cyber-intrusion detection capabilities in complex industrial environments.

Further industrial application anomaly detection developments have focused on improving unsupervised learning approaches. Jiang et al. (2019) introduced a GAN-based anomaly detection model that learned only from normal data, assigning higher anomaly scores to faults without requiring labeled abnormal samples. Niu, Yu, and Wu (2020) tackled time series anomaly detection using an LSTM-based VAE-GAN, which jointly trained encoder, generator, and discriminator to enhance detection accuracy and speed, addressing mapping challenges in real-time anomaly detection. These approaches collectively advance fault diagnosis and anomaly detection, making industrial AI applications more robust, interpretable, and efficient in handling complex and imbalanced datasets.

**A Comparison of Statistics-Based and AI-Related SPC Methods**
**Fig. 2** provides the publication trend of AI-SPC and T-SPC methods for addressing SPC problems starting in 1950. First, a literature search was conducted in "Google Scholar" website using the keyword "statistical process control." Based on these initial search results, the second set of keywords, "artificial intelligence" or "machine learning," was applied. Publications were divided into different eras, as shown in the X-axis in **Fig**. 2. The Y-axis shows the number of publications. This process was repeated using the keyword "statistical process monitoring." The keyword SPM generated much fewer results than SPC as most researchers in this field still prefer the term "SPC." The number of publications has remained about the same for the last 20 years for SPC, but the use of SPM has been increasing. The number of AI/ML SPC and AI/ML SPM publications has increased noticeably in the last five years. The search results from Google Scholar shows a similar trend. However, the details of Google Scholar are not cited in this work.

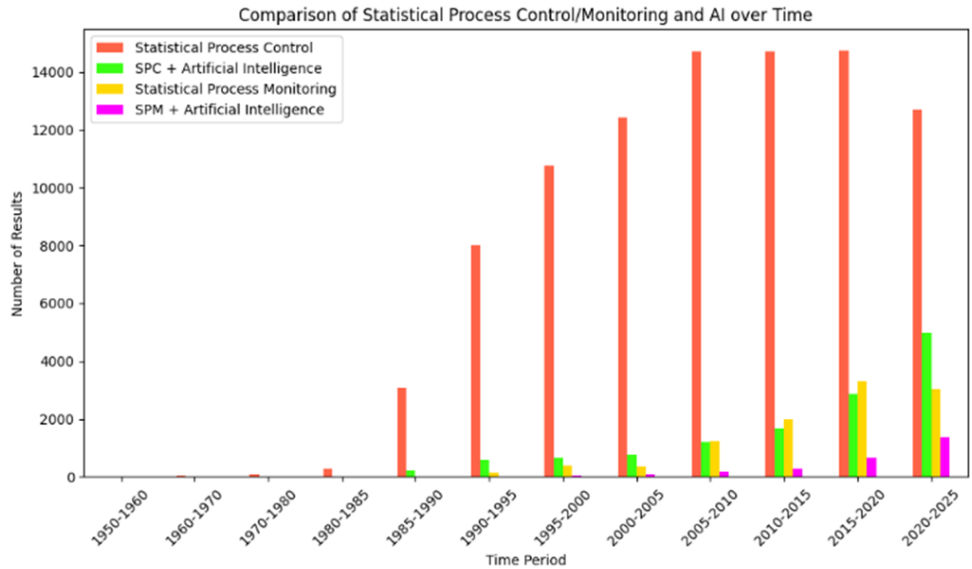

*Fig. 2. Numbers of Publications According to Keywords: SPC, SPM, SPC+ AI, and SPM+AI (from Google Scholar 10/10/2024)*

Out of more than 87,000 SPC/SPM research papers published since 1985, we reviewed more than 260 representative papers in this manuscript. Most SPC applications focus on monitoring univariate and multivariate quality characteristics in one location followed by profile monitoring. Control chart pattern recognitions and multi-stage monitoring do not attract as much attention as the previous two categories. Contrasting the AI-based methods to statistics-based methods, we make the following observations related to the operational aspects of SPM.

***Observation 1: A Collaborative Framework of Applying Statistics-based and AI-based Methods***
Feng et al. (2022) described a collaborative framework of applying both statistics-based and AI-based algorithms for clinical quality improvement applications. Multiple shareholders in a clinical environment include patients, clinicians, model developers, biostatisticians, IT administrators, and regulatory body (i.e. FDA in the US). Since most AI/ML algorithms are data dependent, they may generate adversary predictive results in real-world situations. Worse, certain failure modes may not be detected (Roberts et al. 2014). Currently, trained models must be approved by the FDA and locked (meaning that the models may not be updated without approval.) As a result, the performance of these AI/ML algorithms may gradually diminish. Statistical-based methods can then be used to monitor the outcomes of the AI-based methods. Indeed, due to the regulatory and data challenges, both statistics-based and AI-based approaches can complement each other. Statistics-based methods require much less data while AI-based methods are much more powerful in handling high-dimension and complex relationships but require much large data to function as intended. Statistics-based methods may serve as the guardrail for the AI-based models in a collaborative SPM framework.

*Observation 2: Phase I and Phase II of Control Charting*

In an SPC review paper, Woodall (1997) pointed out that gaps between academic research and industrial practice remain large. Woodall (2000), Stoumbos et al. (2000), and Woodall and Montgomery (1999) also identified some opportunities to narrow these gaps. One remedy is to help users identify appropriate tools and set suitable parameters for the chosen tools in control charting. For example, a typical control chart guideline for control chart setup, also called Phase I of control charting, tells users to choose a "steady state" process period and collect 20 to 25 samples, each with four to five observations, to calculate control limits. Except for the users with extensive statistical training, most users just assume their processes are in steady states without proper analysis for determining independence, constant mean, constant variance, etc. On the other hand, AI-based methods usually require far more initial samples for Phase I setup. Once a control chart is set up, there is no difference between these two strategies in terms of process monitoring from the users' perspective. Note that traditional control charts such as EWMA and CUSUM charts allow individual observation as input, while neural network control charts require multiple input nodes. At first glance, it seems the neural network charts are inferior regarding the sample requirement. However, the configuration of neural network input nodes is very flexible. The sample can be taken from a moving window, which is advanced by one observation at the next sample. Each current sample overlaps with the previous sample except for the current observation. In addition to the moving sample window as input, the NN input can also include the sample mean and standard deviation of the data in the current window. The objective is to detect potential process shifts faster.

*Observation 3: Contrast of Training and Monitoring Philosophies*

Statistics-based SPM (T-SPC where T means traditional) methods focus exclusively on modeling the in-control population. The SPM operation is often viewed as a sequence of hypothesis tests where the null hypothesis $H_0$: the process is in control. If any sample deviates from this hypothesis, the process is deemed out of control. On the other hand, AI/ML-based (AI-SPC) methods such as SVM and ANN often adopt a supervised classification approach that requires the knowledge of both in-control and out-of-control data sets. Usually the AI-SPC methods require balanced numbers of in-control and various types of abnormal data for satisfactory training results. For example, the ANN outputs for detecting mean shifts consists of four categories: no shift (or in control), small mean shift, medium mean shift, and large mean shift. In this case, out of the 4000 training data for this ANN control chart, 1000 observations are generated for no shift, 1-sigma, 2-sigma, and 3-sigma shifts, respectively. When T-SPC control charts are used, it is up to the users to interpret and diagnose the causes of an out-of-control situation. On the other hand, the AI-SPC methods provide more information in terms of diagnoses since the output layers of an ANN control chart can reflect the process status, such as in-control or various out-of-control scenarios. As discussed in observation 1, we favor the collaborative approach that T-SPC methods can be implemented mainly for detection tasks while the AI-SPC algorithms can be used for diagnoses and retrospective analyses of entire data streams. As envisioned by Li and Chen (2021), AI-based methods have great potential for monitoring one person's health data stream from birth and death to achieve preventive personal-based health care.

*Observation 4. Measuring Chart Performance*

We did not find literature that addresses the differences between AI-SPC and T-SPC in terms of chart performance. A satisfactory SPM tool would balance detection speed (measured by type II error) and false alarm rate (type I error). SPM research community for traditional SPC often adopts two metrics: average run length (ARL) in in-control ARL value (i.e., $ARL_0$) and out-of-control $ARL_1$ values for SPM tool evaluation. When comparing various control charts, a common practice is to set $ARL_0$ at a fixed number (e.g., 200 or 370) and then compare various AR1 values. The control charts with small $ARL_1$ values are favorable. On the other hand, the performance of most AI-SPC methods is measured by a different set of metrics, such as accuracy, precision, sensitivity, recall, F1-score, ROC curve, and AUC (area under the ROC curve). A confusion matrix may be the best tool to reconcile the differences in that false negatives (FNR) are Type II errors and false positives (FPR) are Type I errors, assuming that positive means the detecting a shift (i.e., the process is out of control and negative means no detection (i.e., the process is in control).

Unfortunately, most AI-SPC papers do not convert the AI/ML metrics into the ARL values that the traditional SPC research community is often accustomed to. To put this into perspective, a ARL0 of 370 has a type I error of 0.0027 via Normal distribution assumption. It is rare to find a AI-SPC control chart with this property. The metric accuracy is the ratio between the number of correct predictions and the total number of predictions. Therefore, accuracy is a measure of both Type I and Type II errors because a correct prediction avoids making both mistakes. On the other hand, precision is a measure of false positives (FPR, i.e., Type II errors) and recall (or sensitivity) is a measure of false negatives (i.e., Type I errors). Traditional metrics for control chart performance would first fix the in-control ARL (or $ARL_0$) and then generate various out-of-control ARL values from different magnitudes of shifts. If the run length distribution is generated, then the control chart comparison, as shown in **Fig**. 3a is similar to the AOC comparison, as shown in **Fig**. 3b. Since the Type I error (or FPR) is fixed in **Fig**. 3a, we would prefer a control chart whose curve is lower than the other charts due to smaller out-of-control ARL values. On the other hand, in **Fig**. 3b, the preferred method has the largest AUC (area under curve) value.

Many AI-SPC methods may not compete with T-SPC control charts in terms of ARL values if the only decision point is when a sample is deemed out of control at the first sign of trouble. In the case of T-SPC control charts, it is when a point plots outside the control limits. AI-SPC algorithms often use classification methods for process monitoring. When a sample output falls into the small process shift category, the process may be still in control. Therefore, delaying the decision until the next sample may improve the false alarm rates (i.e., decreased Type I errors) without sacrificing out-of-control detection rates (i.e., slightly larger Type II errors). This practice is called the two-sample-in-a-row rule. Finally, T-SPC and AI-SPC methods do not have to compete against each other. When a new process starts, and process data is scarce, T-SPC methods should be implemented. As the process gains traction and more data becomes available, AI-SPC models can be trained to provide more diagnostic information. At this point, T-SPC remains useful for providing guardrails as discussed in Observation 1, while AI-SPC offers a more in-depth analysis at both local and system levels. More details on system monitoring will be discussed in the future direction sections.

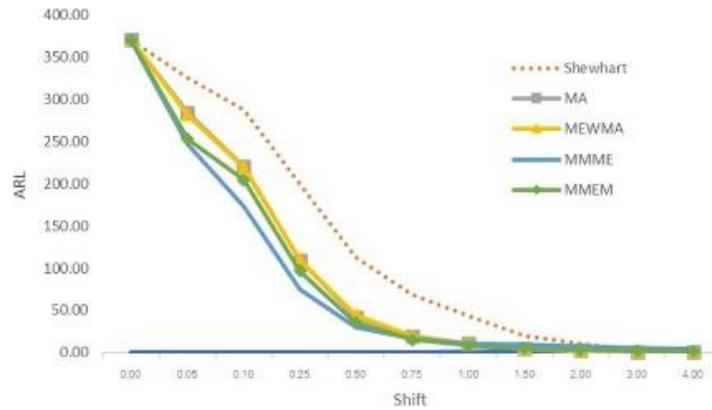

*Fig. 3a. Average run length (ARL) curves of Shewhart, MA, MEWMA, MMME and MMEM control chart for Exponential distribution* (Talordphop, Sukparungsee, and Areepong 2022)

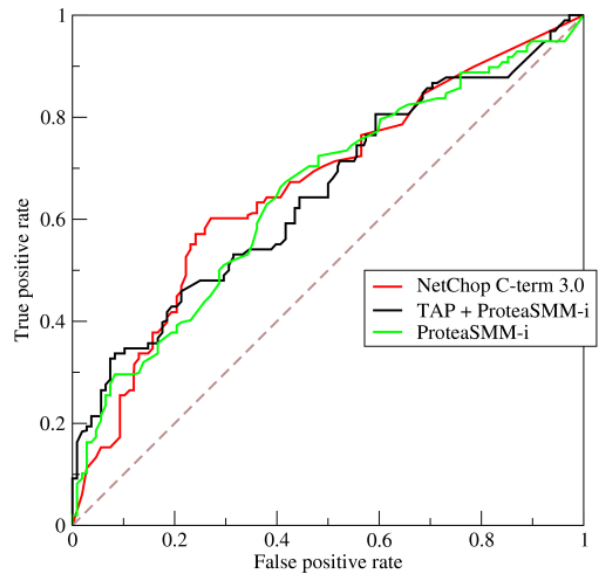

*Fig. 3b. Receiver Operating Characteristic (ROC) curve (True Positive Rate TPR against False Positive Rate FPR)* Wikipedia, The Free Encyclopedia (2025).

    The current performance measures for both T-SPC and AI-SPC methods are primarily focused on efficiency, evaluating how quickly a method can detect deviations while minimizing false alarms. This approach is well suited for most applications. However, as AI-driven methods become increasingly integrated into system decision-making—particularly in healthcare—a broader set of performance measures must be considered, including reliability, interpretability, ethics, and bias.

    For instance, a medical professional relying on AI to monitor patient health would prioritize a system that delivers consistent, unbiased recommendations while adhering to ethical guidelines. Additionally, they would need clear explanations for the AI's suggested treatment procedures. To address these challenges, the SPC research community must expand its evaluation criteria beyond traditional performance metrics.

# Future Directions of AI Applied to SPC

With the advent of generative AI modeling, we articulate two future directions in that AI has great potential to advance existing SPC/SPM research and practices. First, AI-enabled SPC/SPM can be applied to not just one location but the entire system. An example is semiconductor production, where the primary quality character is the yield rate. T-SPC methods on multiple-stage processes as reviewed earlier are limited by the modeling structure due to the system's complexity. It is next to impossible to model all stages in a complex system in mathematical equations (e.g., the state-space models reviewed earlier). We will discuss a generative AI technique called Transformer that has proven quite valuable for modeling spatial and temporal data. The Transformer mechanism is the catalyst for solving large language model problems, leading to the breakthrough in Chat-GPT. Second, existing T-SPC or AI-SPC methods focus on process monitoring only, with most work limited to fault detection or fault diagnosis. The control actions often not addressed are left to process engineers. By fault diagnosis, we mean the location and magnitude of the responsible process parameters contributing to the fault. The ultimate goal of SMPC (**S**mart **P**rocess **C**ontrol) is to restore a faulty local process to its original state autonomously. Most importantly, fault prevention is more desirable. In this SMPC framework, a preventive SPC algorithm, rather than the reaction-oriented SPM approach, may provide warning signals triggering preventive actions so that the out-of-control situations may be adverted. In a semiconductor fabrication process, the preventive SPC may spot deviations in one stage of a cascading production process rather than an attribute control chart on yield used at the end of the production process. In a personal health monitoring application, data from annual check-ups may trigger preventive interventions when all data related to this patient is considered. For example, the existing protocol for rehab treatment of an elderly patient is when this patient suffers a fall. A preventive intervention is to send this patient for muscle strength exercises so that falls may be prevented. A preventive SMPC may identify this patient's walking patterns and fall risks (Malmir et al. 2019). In the following sections, we will discuss potential AI modeling approaches that may be adopted for the envisioned SMPC. Challenges of AI model building will also be addressed.

**Future Direction 1. Generative Neural Network Model for Complex Systems**

The use of Generative Neural Network (GNN) models in Statistical Process Monitoring (SPM) provides a powerful way to handle the challenges of modern industrial systems. These systems often deal with large amounts of data, complex behaviors, and nonlinear interactions. GNNs can learn detailed patterns and create synthetic data, helping improve process monitoring and control. Generative models like Variational Autoencoders (VAE), Generative Adversarial Networks (GANs), and Neural Ordinary Differential Equations (NODEs) are good at understanding data patterns and relationships (Kingma 2013, Lyu et al. 2019). They can create realistic data to simulate different scenarios, find problem areas, and predict potential failures. Using GNNs in SPM goes beyond traditional methods, offering more flexible and responsive control systems.

   Generative models help detect anomalies by learning normal system behaviors and identifying unusual changes that signal problems. For example, GANs can spot abnormalities by comparing real-time data with synthetic data generated under normal conditions (Losi et al. 2025). GNNs can also simulate scenarios to optimize process settings, reducing variation and improving quality. This is especially helpful in systems where many variables interact in complex ways (Chung, Shen, and Kong 2024). VAEs and NODEs can model time-based patterns in machine behaviors, predicting issues before they happen, which reduces downtime and increases efficiency (Mohammadi, Mahmoud, and Elbestawi 2021).

Detecting defects in semiconductor manufacturing is critical for improving yield and wafer quality. Kim et al. (2021) emphasized the importance of generative models, particularly GANs, in learning normal system behaviors and identifying anomalies that signal defects. For instance, GANs compared real-time data with synthetic data generated under normal conditions to spot abnormalities effectively. This approach has proven instrumental in advancing defect detection methods in semiconductor processes.

GNNs handle large datasets by extracting important features and simplifying data without losing key information. Alaa et al. (2022) can easily scale the model to manage large, complex systems with many variables (Goyal and Mahmoud 2024). GNNs adjust to changes in system dynamics, ensuring they respond in real time. Physics-informed models combine domain knowledge with data-driven insights to make accurate and understandable predictions.

Despite their benefits, using GNNs in SPM comes with challenges. Training these models requires significant computational resources, especially for large datasets. High-quality data is often needed but can be hard to obtain. Additionally, the "black-box" nature of GNNs makes it difficult to understand how they make decisions, which can be a problem in critical applications (Wang et al. 2017).

To overcome this challenge, synthetic training data should be generated from a small amount of sample data representing each fault category (Moreno-Barea, Jerez, and Franc 2020; Figueira and Vaz 2022). From the above reviews, GAN is a good tool to generate synthetic data. However, GAN models are difficult to train and very sensitive to hyperparameter variation (Bowles et al. 2018; Alaa et al. 2022) . Recently, diffusion models gain traction because the diffusion framework offers stable training, high sample diversity although it usually requires more computing power and training time compared to a GAN model (Dhariwal and Nichol 2021; Wang et al. 2023).

Future work should focus on combining GNNs with traditional methods to improve clarity and performance, reducing computational requirements for real-time use, and developing models that are easier to interpret and trust. Integrating GNNs with IoT sensors and edge computing could create decentralized and scalable systems for process control. The use of GNNs in SPM is still growing but has the potential to revolutionize process control. By combining traditional statistical methods with advanced AI techniques, industries can achieve smarter, more adaptable, and more efficient process monitoring and control.

Another potential breakthrough in SPC applications may involve integrating transformer architectures from generative AI models. These models are commonly used in large language models (LLMs) originally designed for tasks such as translating English to German (Vaswani 2017). At the core of a transformer is the self-attention mechanism, which enables the model to process and relate different parts of an input sequence efficiently. Given a sequence of tokens (e.g., words), a trained LLM predicts the next token based on learned probabilities. If the token represents a word, the model generates a probability distribution over all possible tokens within the application's vocabulary. The token with the highest probability is appended to the sequence, and this process continues iteratively until the task is complete.

Dosovitskiy et al. (2020) extended the self-attention mechanism to image classification, such as "An Image is worth 16x16 words", where an image is divided into 16x16 stamps, then strung together like a sentence of 16x16 words. The recent introduction of Chat-GPT 4o (OpenAI et al. 2024) has further incorporated images, audio, and videos in addition to text as both inputs and outputs. Google has also announced its Large Multi-modal Model (LMM) in its AI Gemini release (Pichai and Hassabis 2023). Chen and Ho (2022) demonstrated a model called MM-ViT, combining both compressed video and audio inputs for action classification.

These recent developments in generative AI modeling have shown the SPC research community a path to predict the next stage of the process, assuming a proper token can be designed. We believe the encoding and decoding mechanisms described in the previous paragraphs may hold the key to the SPC token design in that the output of the encoding stage is a feature map. Features may be generated from this map that achieve dimension reduction. Infinite combinations of input variable states in multiple process stages need to be reduced in a finite, manageable token Scopus. To put this in perspective, the Chat-GPT 4o LMM model has a maximum output token limit of 4,096. The total number of input and output tokens is limited to 128,000 in a single interaction (Dizon 2024). Assuming that the data for the process parameter vector $X_t^i = (x_1, x_2, \ldots)_t^i$. At stage i and time t, we can gather all process vectors that generate a semi-finished or finished product into a string of vectors. This is very similar to the joint features idea by Jiang, Yan, and Huang (2019) in the review of the multi-stage QE applications section. The timestamps for all collected process vectors indicate the states under which this semi-finished or finished product is produced.  Based on this string at the current time, the task is to predict the probability of the most likely token for the next stage. In this framework, each stage contains a set of tokens that reflect the most likely outcomes from that stage. The collection of all tokens forms the token Scopus. When the predicted token deviates from the actual outcome, the difference will provide a base for predicting the likelihood that a process is trending out of control or staying in control. The main challenge for this framework is the data requirement for model training. In the case of semiconductor manufacturing, the benefit of modeling the entire production stages certainly outweighs the data collection challenge. This data-driven framework alleviates the modeling barriers of traditional state-space method applied to complex systems.

**Future Direction 2. Generative AI Model for True Process Control**
The transform architecture discussed in future direction 1 can also be extended to enable process control. We now provide an example of a food production ecosystem as shown in **Fig**. 4 so that the proposed **SPC (Smart Process Control)** vision can contrast with the existing SPC (Statistical Process Control) practices. While the concept of applying SPC towards process monitoring is sound, the existing SPC techniques do not address process control at all. The AI framework illustrated in **Fig**. 4 represents a Controlled Environment Agriculture (CEA) system utilizing hydroponics, capable of making mid-growing-cycle adjustments. A generative AI model, leveraging the self-attention mechanism, is employed to capture the temporal relationships between input variables and optimize CEA adjustments. Specifically, the inputs include aerial and root images $I_t$ and current setting of the CEA variable vector ***x=(x₁, x₂, …)***. Examples of the elements in vector ***x*** include temperature, humidity, $CO_2$ levels, airflow and ventilation, light management (light intensity, light spectrum, and photoperiod), irrigation, and nutrient water feeding. The system output for control purposes are the changes of CEA variable vector, **Δ*x*** at time t***.*** We divide a growing cycle into *n* stages. For example, lettuce typically goes through five main growing stages in seed germination, seedling, vegetative growth, maturity & head formation, and harvesting that lasts about 45-70 days depending on variety. We can establish a daily control system or implement a control protocol tailored to the key growth stages. Time-lapse aerial and root images $I_t$ within a stage are then strung into a video. CEA parameters associated with each stage are collected in a vector $x_t$.

Through the self-attention mechanism, a score is generated for each element that feeds into a sequence of transformation operations that determine which CEA parameters would contribute the most to improving the outcome. Specifically, any input vectors feeding into a self-attention mechanism generate three components: Q, K, V via dot product operations, respectively. Then, self-attention scores are computed via another dot product operation. After a SoftMax operation normalizes each column of all dot product vectors, they are collected column-wise in the output matrix O. These output vectors in O from the self-attention box (e.g., the purple box in **Fig**. 4) feed into a fully connected network (FCN) to generate **Δx** at time **t**. Note that the weight parameters self-attention box (called a transformer) and FCN must be trained. All other matrices are provided. In general, a generative model requires more data to train than a CNN, a RNN (recurrent neural network), or a hybrid of CNN/RNN model. Since CNN does not carry memory, RNN architect is a way to string a series of CNN models where outputs of a prior CNN feed into its adjacent CNN to bring the memory required to solve a problem. The proposed generative model is more flexible in connecting all input vectors. In this case, the input vectors carry both the crop images and CEA parameters setting over time. This flexibility allows the early results to possibly affect later outcomes without making prior assumptions. The model performance depends solely on the training data rather than model assumptions!

The main challenge of any AI framework including this CEA application is the requirement of big data. The data required for training generative AI models is very demanding. Many growing cycles are required to generate a large enough training data set. It took Tesla many years to collect a large fleet of cars collecting millions of miles of driving data to perfect its FSD application. Chang et al. (2024) propose a digital twin that collects CEA data from many vertical farming (VF) farmers. Similar to millions of Tesla cars collecting driving data, thousands of VF operations may provide enough data to make the proposed generative AI training possible. During the initial implementation stage, GAN and diffusion models can be used to generate synthetic data based on limited CEA data. Another barrier to training a large model with generative AI models is computational power. This computational restriction will ease over time as the recent AI infrastructure investments from various tech giants such as Microsoft, Google, and Meta will provide ample computational resources. The computational challenges of AI model training may no longer be a limiting factor. We predict that the bottleneck of the proposed CEA models is not the computational hardware but the data available for training.

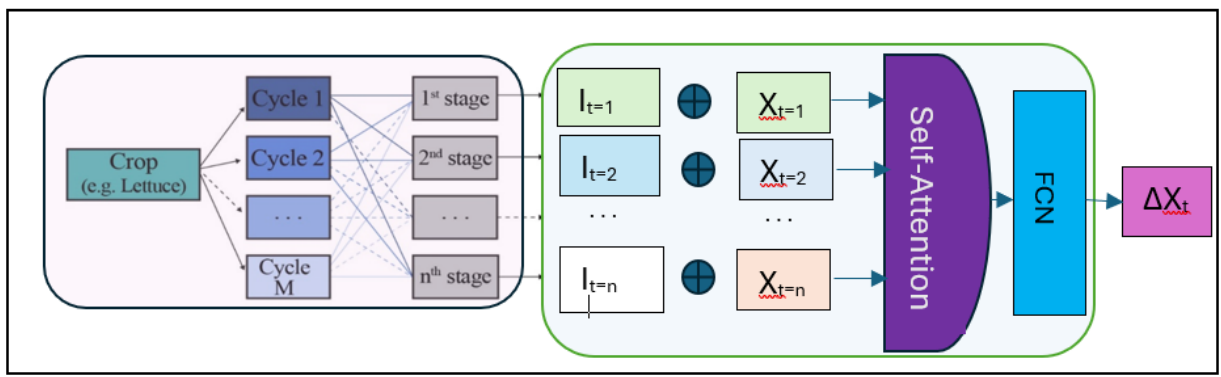

*Fig. 4. A Self-Attention Mechanism in DNN for Mid-Cycle Adjustments (Chang et al. 2024)*

## Conclusions

SPC has evolved over a century, focusing on monitoring univariate, multivariate, and profile-based quality characteristics. This manuscript offers a comprehensive review of the intersection between traditional Statistical Process Control (SPC) methods and Artificial Intelligence (AI) technologies for process monitoring, emphasizing their applications in various industrial settings. It highlights how traditional SPC methods, rooted in statistical approaches, are now complemented or replaced by AI techniques such as neural networks, convolutional neural networks, and generative models for more dynamic, high-dimensional, and complex monitoring needs. We also present insights into how traditional and AI-based SPC methods can work together synergistically to complement each other. To fully evaluate the performance of AI-SPC methods, additional metrics are required to assess not only detection speed, accuracy, sensitivity, and specificity but also reliability, interpretability, ethics, and bias.

AI methods (machine learning included) have shown tremendous potential in process monitoring applications in recent years. This review categorizes AI applications in SPC into areas like classification, anomaly detection, and generative AI for proactive fault detection. Methods such as GANs, GNNs, and transformers represent a promising future for handling high-dimensional, multistage processes. AI advancements, including generative AI and multimodal models, are projected to evolve SPC into Smart Process Control (SMPC), enabling predictive and autonomous process corrections.

Karwowski et al. (2025) listed eight grand challenges in industrial and systems engineering. Two grand challenges are related to AI, specifically, (1) the AI for business and personal use: decision making and system design and operations, and (2) system integration and operations: humans, automation, and AI. As individual AI applications (e.g. anomaly detection AI agents) are trained and deployed throughout a system. It becomes more crucial for an integrated AI implementation strategy to harvest fruits but avoid the pitfalls. The path forward is either a centralized AI system to control all components and stages directly or decentralized AI agents coordinated by a master AI decision-making agent. It is certainly possible to have a hybrid variant of an ensemble system. We are just at the start of this AI evolution that will certainly impact the SPM research and practices.

In conclusion, the integration of AI into SPC marks a paradigm shift, enabling more robust, efficient, and intelligent process monitoring systems. However, challenges like computational demands, interpretability, and the need for high-quality data remain key hurdles to address in future research and industrial applications.